\documentclass[aps,prd,twocolumn,showpacs,superscriptaddress,nofootinbib,floatfix]{revtex4-1}  
\usepackage{graphicx}  
\usepackage{dcolumn}   
\usepackage{bm}        
\usepackage{comment}
\usepackage{amssymb}   
\usepackage{amsmath}
\usepackage{tabularx}
\usepackage{hyperref}
\usepackage{tabulary}
\usepackage[usenames,dvipsnames]{color}
\usepackage[normalem]{ulem}

\begin{document}

\preprint{APS/123-QED}

  \title{Dynamical friction in dark matter spikes: corrections to Chandrasekhar's formula
  }
   \author{Fani Dosopoulou}
   \email{dosopoulouf@cardiff.ac.uk}
    \affiliation{School of Physics and Astronomy, Cardiff University, Cardiff, CF24 3AA, United Kingdom}
\affiliation{Princeton Center for Theoretical Science, Princeton University, Princeton, NJ 08544, USA} 
\affiliation{Department of Astrophysical Sciences, Princeton University, Princeton, NJ 08544, USA}
\date{\today}

\begin{abstract}
 We consider the intermediate mass-ratio inspiral
of a stellar-mass compact object with an intermediate-mass black hole that is surrounded by a dark matter density spike. The interaction of the inspiraling black hole with the dark matter  particles in the spike leads to dynamical friction. This can 
 alter the dynamics of the black hole binary, leaving an imprint on the gravitational wave signal.
Previous calculations did not include in the evaluation of the dynamical friction coefficient the contribution from particles that move faster than the inspiraling black hole. This term is neglected in the standard Chandrasekhar treatment where only slower moving particles contribute to the decelerating drag. Here, we 
demonstrate that  dynamical friction produced by the fast moving particles can have a significant
effect on the evolution of a massive binary within a dark matter spike.
For a density profile $\rho\propto r^{-\gamma}$ with $\gamma\lesssim 1$, the dephasing of the gravitational waveform can be several orders of magnitude larger than estimated using the standard treatment.
As $\gamma$ approaches $0.5$ the error becomes arbitrarily large.
Finally, we show that dynamical friction tends to make the orbit more eccentric for any $\gamma < 1.8$. However, energy loss by gravitational wave radiation is expected to dominate the inspiral, leading to orbital circularization in most cases.
\end{abstract}
\maketitle


\section{Introduction}
\label{sec:intro}
The inspiral of an intermediate mass black hole (IMBH)  and a solar mass type object will be observable by space-based gravitational wave (GW) detectors such as the Laser Interferometer Space Antenna (LISA)\cite{AmaroSeoane:2012km,2017arXiv170200786A,2023LRR....26....2A,Barausse:2014tra}.
Around these IMBHs, a dark matter (DM) halo could grow adiabatically into a DM spike \cite{Gondolo:1999ef,Sadeghian:2013laa}. These spikes have extremely high densities and can leave an imprint on the GW signal emitted by the binary by modifying its orbital evolution.
This opens up the possibility to infer the existence and the properties of the DM spike from  measuring its
impact on the GW signal \cite{Eda:2013gg,Eda:2014kra}.

The predicted effect is a dephasing of the gravitational waveform due to dynamical friction the secondary object experiences
while passing through the mini spike. This
decelerates the secondary object and results in a faster inspiral,
which would be observable in the phase evolution of the
GW signal \cite[e.g.,][]{Yue:2017iwc,Macedo:2013qea,Yue:2019ndw,Cardoso:2019rou,2020PhRvD.102h3006K,2023PhRvD.107h3003B}.

The dynamical friction force on the massive binary is calculated in the literature following the standard formulation of Chandrasekhar \cite{Chandrasekhar1943a,Chandrasekhar1943b}. This has been also somewhat modified to include the back reaction of the DM spike to the binary motion \cite{2020PhRvD.102h3006K}, which is expected to flatten the inner cusp \cite[e.g.,][]{Merritt:2006mt,2021ApJ...922...40D}; the inclusion of  relativistic terms in the treatment of the orbital dynamics and distribution of DM \cite{2022PhRvD.106d4027S,2022arXiv220705728C};
and DM accretion into the small compact object \cite{Macedo:2013qea,Yue:2019ndw}.

The calculation of the coefficient of dynamical friction is done by assuming that only DM particles that move slower than the inspiraling object contribute to the decelerating force. This corresponds to  Chandrasekhar's result that stars moving faster than an inspiraling object have a negligible contribution to dynamical friction \cite{Chandrasekhar1943a,BinneyAndTremaine}.
However, this approximation has been shown to break down
 when the gravitational potential around the binary is nearly Keplerian, as it is the case under consideration. \cite{2012ApJ...745...83A} and \cite{2017ApJ...840...31D} showed that in a cusp where the density falls off more slowly than $\rho \propto r^{-1}$, the contribution of the fast-moving particles to the fictional force becomes dominant and cannot be neglected.

In this work, we present a proof-of-concept analysis of the evolution of a massive binary in a DM cusp. For the first time, we include the dynamical friction force due to DM particles moving faster than an inspiraling black hole (BH). 
We compare to the standard treatment, and  quantify the error made when this term is neglected.

We begin in Section \S\ref{formulation} by introducing our formulation, including the orbit-averaged equations that 
describe the binary evolution due to dynamical friction and  energy loss due to GW radiation.
In \S\ref{effectorb} we explore the effect of the additional dynamical friction term on the orbital decay time of the binary, the evolution of its eccentricity, and study their dependence on the density profile slope of the DM spike.
Finally, in \S\ref{dephasing} we study the effect on the dephasing of the GW signal emitted by the binary. In \S\ref{conc} 
we summarize our main results and conclude.

\section{Formulation}\label{formulation}

The general formula for the dynamical friction force a BH experiences during its inspiral inside a DM cusp is \cite{2012ApJ...745...83A,2017ApJ...840...31D}
\begin{align} 
\begin{split}
 \pmb{F}_{df}\approx&-4\pi G^{2} m \rho(r) \frac{\pmb{\upsilon}}{\upsilon^{3}}
\times \left\lbrace \ln{\Lambda} \int_{0}^{\upsilon}d\upsilon_{\rm DM} 4\pi f(\upsilon_{\rm DM})\upsilon_{\rm DM}^{2}\right.\\
 &\left. +\int_{\upsilon}^{\upsilon_{\rm esc}}d\upsilon_{\rm DM}4\pi
 f(\upsilon_{\rm DM})\upsilon_{\rm DM}^{2}\left[\ln\left({\frac{\upsilon_{\rm DM}+\upsilon}{\upsilon_{\rm DM}-\upsilon}}\right)-2\frac{\upsilon}{\upsilon_{\rm DM}}\right] \right\rbrace\end{split}\label{df}
\end{align}
   {where $\pmb{\upsilon}$ is the velocity of the inspiraling BH and $m$ its mass; $\rho(r)$ is the local density of the DM particles,  $f(v)=(1/n_{M})d^{6}N/d^{3}xd^{3}v$ with $n_{M}=d^{3}N/d^{3}x$ is the normalized phase distribution (assumed to be isotropic),} and $v_{\rm esc}$ is the escape velocity.
The quantity $\ln{\Lambda}$ is the Coulomb logarithm defined as
\begin{equation}\label{coulomb}
\ln{\Lambda}=\ln\left(\frac{b_{\rm max}}{b_{\rm min}}\right)\approx \ln\left(\frac{b_{\rm max}v_{c}^{2}}{Gm}\right)
\end{equation}
where $b_{\rm max}$ and $b_{\rm min}$ are the maximum and minimum impact parameters respectively and $v^2_{c} =GM_{\bullet}/r$ the circular velocity around the IMBH of mass $M_{\bullet}$. The first integral term in the RHS of Eq.\ (\ref{df})  represents the decelerating drag due to DM particles moving slower than the infalling BH.
The second integral term instead,  represents the contribution from particles moving {\it faster} than the BH. This latter
term is often neglected because it is typically a factor $\sim \ln \Lambda$ smaller than the former term. But, we will show below that under some specific conditions about the surrounding DM density profile and kinematics, the fast particle contribution becomes dominant.

 We can rewrite the dynamical friction force as
\begin{align}
\bold{F}_{df} &= \epsilon (r,v) \frac{\pmb{v}}{v^{3}}\label{df2}
\end{align}
where we defined 
\begin{align}
\epsilon (r,v) &= -4\pi G^{2}\rho(r)m \left[ \ln{\Lambda}\:\: \alpha(v) + \beta (v) + \delta (v) \right]\label{int2}\\
\alpha (v) &= 4\pi\int_{0}^{v} f(\upsilon_{\rm DM}) \upsilon_{\rm DM}^{2} \, d\upsilon_{\rm DM}\label{int3}\\
\beta (v) &= 4\pi\int_{v}^{v_{\rm esc}} f(\upsilon_{\rm DM}) \upsilon_{\rm DM}^{2} \left[\ln\left({\frac{\upsilon_{\rm DM}+\upsilon}{\upsilon_{\rm DM}-\upsilon}}\right)\right] d\upsilon_{\rm DM}\label{int4}\\
\delta (v) &= 4\pi v \int_{v}^{v_{\rm esc}} f(\upsilon_{\rm DM}) (-2\upsilon_{\rm DM})\, d\upsilon_{\rm DM}.\label{int5}
\end{align}

The {osculating orbital element time-evolution equations of the inspiraling BH inside the DM cusp surrounding the IMBH due to dynamical friction} are \cite{2017ApJ...840...31D}
\begin{align} 
\frac{da}{dt}&=\frac{2\epsilon (r,v)}{n^{3}a^{2}}\frac{(1-e^{2})^{1/2}}{(1+e^{2}+2e\cos{f})^{1/2}}\label{da}\\
\frac{de}{dt}&=\frac{2\epsilon (r,v)}{n^{3}a^{3}}(1-e^{2})^{3/2}\frac{e+\cos{f}}{(1+e^{2}+2e\cos{f})^{3/2}}\label{de}\\
\frac{d\omega}{dt}&=\frac{2\epsilon (r,v)}{n^{3}a^{3}}\frac{(1-e^{2})^{3/2}}{(1+e^{2}+2e\cos{f})^{3/2}}\frac{\sin f}{e}\label{domega}\\
\frac{df}{dt}&=\frac{n (1+e\cos f)^{2}}{(1-e^{2})^{3/2}}-\frac{d\omega}{dt}-\cos i \frac{d\Omega}{dt}\label{dsigma}.
\end{align}
where $a$ is the semi-major axis, $e$ the eccentricity, $\omega$ the argument of periapsis, $\Omega$ the longitude of the ascending node and $f$ the true anomaly. $M=m+M_{\bullet}$ is the total binary mass and $n=2\pi/T=(GM/a^{3})^{1/2}$ the orbital angular frequency,
   {We note that the perturbation of the binary orbit in this case is caused only by dynamical friction. Since there is no vertical component of the dynamical friction force to the orbital plane, the inclination $i$ and the longitude of the ascending node $\Omega$ remains constant in the absence of other perturbing forces and thus $\dot{i}=\dot{\Omega}=0$.   Dynamical friction, however, induces a precession to the argument of periapsis, $\omega$,  and changes the orbital  semi-major axis $a$ and eccentricity $e$. This latter can either increase or decrease depending on the density profile slope of the DM spike adopted.}

Since the time evolution of the orbital elements occur over many orbits, in order to determine the time evolution of the massive binary orbit we can orbit-average the above equations.    {In order to orbit-average a quantity along the orbit, we need to know how the true anomaly is changing over time. This is described by Eq. (\ref{dsigma}), which shows that apart from the unperturbed Keplerian evolution described by the first term on the RHS of Eq. (\ref{dsigma}), the true anomaly can also evolve due to possible precessions and specifically the periapsis precession $\dot{\omega}$ and the longitude of the ascending node precession $\dot{\Omega}$. Given that due to dynamical friction we have $\dot{\omega}<< 1$ and that $\dot{\Omega}=0$, we can compute the secular evolution of the orbital elements neglecting the second and third term in Eq. (\ref{dsigma}) and use}

\begin{equation}
df=n \frac{(1+e\cos{f})^{2}}{(1-e^{2})^{3/2}}dt.
\end{equation}

Under these considerations, the secular time-evolution equations of the binary orbit are

\begin{align}
\left\langle\: \frac{da}{dt} \:\right\rangle_{\rm DF}=&\frac{(1-e^{2})^{2}}{\pi n^{3}a^{2}}
\int_{0}^{2\pi}\:\frac{(1+e\cos f)^{-2}\epsilon (r,v)}{(1+e^{2}+2e\cos{f})^{1/2}}df\label{avinta}\\
\nonumber \left\langle\: \frac{de}{dt} \:\right\rangle_{\rm DF}=&\frac{(1-e^{2})^{3}}{\pi n^{3}a^{3}}\\
&\times \int_{0}^{2\pi}\:\frac{(e+\cos f)\epsilon (r,v)}{(1+e^{2}+2e\cos{f})^{3/2}(1+e\cos f)^{2}}df,\label{avinte}\\
\left\langle\: \frac{d\omega}{dt} \:\right\rangle_{\rm DF}=& 0   \ .
\end{align}
Thus, over many orbits dynamical friction will change the semi-major axis and eccentricity of the orbit but on average it will cause no in-plane precession of the orbit.

Below a certain distance between the two BHs, energy loss by GWs becomes important and needs to be added to the dynamical friction effect. This is certainly the case for a massive binary in the LISA frequency band $0.1\rm mHz\ - 1 \rm Hz$.
The orbital evolution due to GW energy loss is \cite{1964PhRv..136.1224P}:
\begin{align}
\left\langle\: \frac{da}{dt} \:\right\rangle_{\rm GW}=&
-{64G^3mM_{\bullet}\left(m+M_{\bullet}\right)\over 5c^5a^3(1-e^{2})^{7/2}}
f_1(e)
\\
\left\langle\: \frac{de}{dt} \:\right\rangle_{\rm GW}=&
-{304G^3mM_{\bullet}\left(m+M_{\bullet}\right)\over 15c^5a^4(1-e^{2})^{5/2}}f_2(e)
\end{align}
where the eccentricity dependent terms are $f_1=1+{73 \over 24} e^2+{37\over 96}e^4$ and $f_2=1+{121 \over 204} e^4$.

Finally, we obtain the evolution of the inspiralling BH orbit  by integrating the following coupled set of first order differential equations:
\begin{align}\label{eqa}
\left\langle\: \frac{da}{dt} \:\right\rangle=\left\langle\: \frac{da}{dt} \:\right\rangle_{\rm DF}+\left\langle\: \frac{da}{dt} \:\right\rangle_{\rm GW}
\\\label{eqe}
\left\langle\: \frac{de}{dt} \:\right\rangle=\left\langle\: \frac{de}{dt} \:\right\rangle_{\rm DF}+\left\langle\: \frac{de}{dt} \:\right\rangle_{\rm GW}.
\end{align}

\section{Effect on the binary orbit}\label{effectorb}
We consider here the simple case in which the DM cusp is a  power law profile
\begin{equation}
\rho(r)\propto r^{-\gamma}\ .
\end{equation}

\begin{figure}\hspace*{-1.5cm}
\includegraphics[width=0.45\textwidth,angle =270]{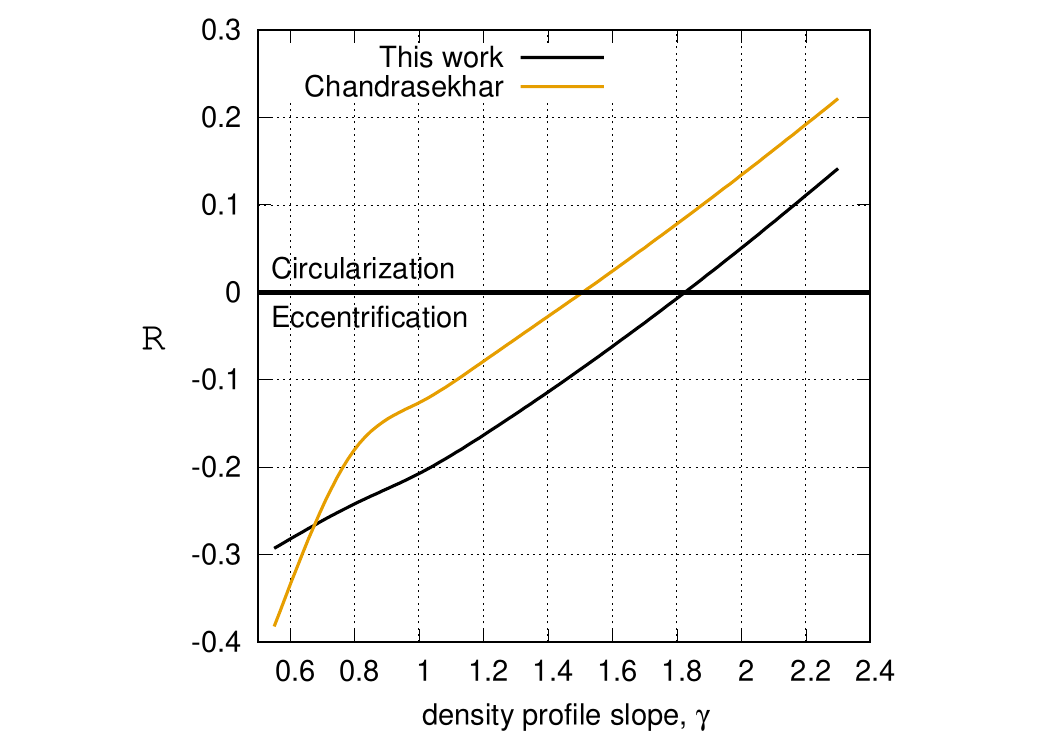}
\caption{Evolution of eccentricity due to dynamical friction, including the contribution form the fast moving particles. In cusps with $\gamma >1.8$, dynamical firction causes the orbit to circularize faster than if it was evolved only due to energy loss by GW radiation. For shallower slopes, instead, the orbit is expected to 
circularize at a slower rate. Note that in the standard treatment
of Chandrasekhar, the transition occurs at $\gamma=1.5$.
In this calculation we did not include the 2.5pN terms that are always dominant and cause the orbit to circularize.
}\label{eccen}
\end{figure}

\begin{figure}\hspace*{-1.5cm}
\includegraphics[width=0.45\textwidth,angle =270]{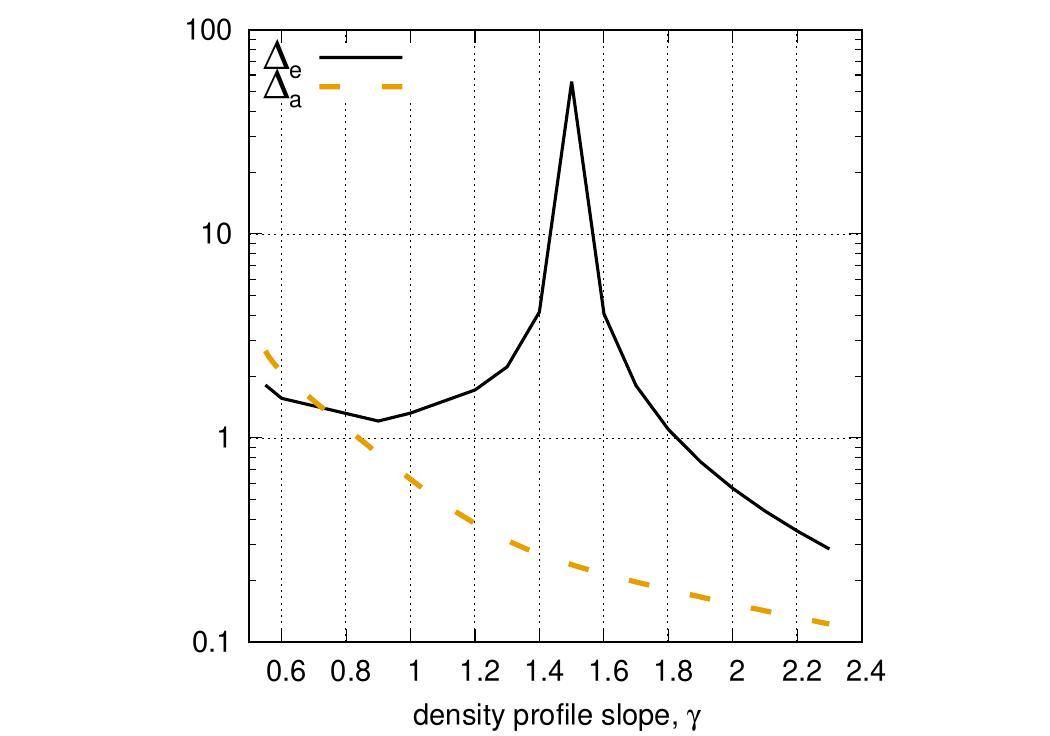}
\caption{   {
Relative error in the dynamical friction evolution timescale made when neglecting the contribution from the fast moving particles. We plot 
$\Delta_a = \frac{\left| T_{\rm C; a} - T_{\rm df; a} \right|}{T_{\rm df; a}}$ and 
$\Delta_e = \frac{\left| T_{\rm C; e} - T_{\rm df; e} \right|}{T_{\rm df; e}}$, as defined in the main text, as a function of the density profile slope. For $\gamma<0.8$, the standard formula predicts an orbital decay time that is more than twice as long than predicted by 
our treatment. Deviations in $e$ is significant for any value of $\gamma$. For $\gamma = 2$, our treatment predicts a circularization
timescale that is about 2 times shorter than the standard
formula.}
}\label{decay}
\end{figure}

Assuming that the gravitational potential $\Phi$ is dominated by the central IMBH and neglecting the effect of the surrounding DM particles we can write $\Phi\approx-G M_{\bullet}/r$. Eddington's formula then uniquely leads to the following distribution function of the 
DM particle velocities \citep{2013degn.book.....M}
\begin{equation}
f(\upsilon_{\rm DM})=\frac{\Gamma\left(\gamma+1\right)}{\Gamma\left(\gamma-\frac{1}{2}\right)}\:\frac{1}{2^{\gamma}\pi^{3/2}v_{c}^{2\gamma}}
\:\left(2v_{c}^{2}-\upsilon_{\rm DM}^{2}\right)^{\gamma-3/2}\label{dis1}
\end{equation}
where the normalization 
corresponds to  unit total number. We employ the distribution function to calculate the effect of dynamical friction on the orbit of the inspiraling BH. 

We consider a binary with primary IMBH mass $M_{\bullet}=1.4 \times 10^3\ M_\odot$, mass ratio $q=m/M_{\bullet}=10^{-3}$, initial $a_0=10^{-8}\rm pc$
and $e_0=0.1$. {    Based on these initial conditions, the binary is intially in the LISA frequency band, $0.1\rm mHz\ - 1 \rm Hz$, with a small but finite eccentricity}. We evaluate the importance of dynamical friction on the evolution of the binary $a$ and $e$, and comment on the effect of the ``non-dominant'' terms.

{    Fig. \ref{eccen} shows the quantity
$
\mathcal{R} = {\frac{1}{e} \left\langle \dot{e} \right\rangle_{\rm DF}}/{\frac{1}{a} \left\langle \dot{a} \right\rangle_{\rm DF}}
$
as a function of $\gamma$; $\mathcal{R}$ represents an approximation of the fractional change in eccentricity in a time 
$a/\left\langle\: \dot{a} \:\right\rangle _{\rm DF}$, i.e., the orbital decay timescale due to dynamical friction. 
For $\mathcal{R}<0$ the orbit becomes more eccentric, for $\mathcal{R}>0$ it circularizes, and for $\mathcal{R}=0$ the eccentricity remains constant.
}
From the value of $ \mathcal{R}$, we expect dynamical friction to have a small effect on the evolution of the binary's eccentricity. The evolution remains dominated by GW energy loss, which leads to the circularization of the binary.  This conclusion is in contrast to what stated in \cite{2021PhRvD.103b3015C}, who finds that generally the binary becomes more eccentric. 
\cite{2022PhRvD.105f3029B} find that by including the relative velocities of the DM particles, dynamical friction tends to circularize the orbit. We agree with the latter authors, but find that dynamical friction tends to circularize the orbit for a smaller range of density profile slopes. Using our more complete formulation, we find that the effect is to circularize the orbit for any $\gamma\gtrsim 1.8$. Instead, the standard Chandrasekhar's treatment 
predicts orbital circularization for any $\gamma\gtrsim 1.5$ \cite{2003ApJ...592..935G}.

{   
Ref.~\cite{2022PhRvD.105f3029B} show that at first order 
$(a/e) (de/da) = \gamma/2$
when dynamical friction is the dominant
form of energy loss and in the limit of slow-moving particles, demonstrating 
 that
DM effects can be observable not just from dephasing but also from the circularization rate. Our results agree with this conclusion, but require a modification of the binary orbit circularization rate due to the fast moving particles.
}


Fig. \ref{decay} further quantifies the deviation from Chandrasekhar's treatment as a function of the power law index of the density profile. We 
{    compute the relative error that one would make by
 using the standard Chandrasekhar's formula compared to our treatment.
This is obtained as
$\Delta_a={\left| T_{\rm C; a}-T_{\rm df; a}\right| \over T_{\rm df; a}}$, and $\Delta_e={\left| T_{\rm C; e}-T_{\rm df; e}\right| \over T_{\rm df; e}}$   
where
\begin{equation}
T_{\rm df;\ a}^{}={ a\over  \left|\left<\dot{a}\right>_{\rm DF}\right|}\ ;
\end{equation}
is the orbital decay timescale, and
\begin{equation}
T_{\rm df;\ e}^{}={ e\over  \left|\left<\dot{e}\right>_{\rm DF}\right|}\ ;
\end{equation}
is the timescale of eccentricity evolution.
$T_{\rm C; a}$ and $T_{\rm C; e}$ are the dynamical friction timescales for orbital decay and eccentricity evolution obtained from the standard Chandrasekhar's formula, respectively.
}

From this analysis, we can see that the contribution from the fast moving particles on the evolution of $a$ becomes more important as $\gamma$ approaches $0.5$ from above. For $\gamma<0.8$, the standard formula predicts an orbital decay time that is more than twice as long than predicted by the
our treatment.
The contribution from the fast moving particles has also an effect on $e$. Deviations in this case remains significant for essentially any value of $\gamma$. For $\gamma=2$,  our treatment predicts a circularization timescale that is about 2 times shorter than the standard formula.

\begin{figure}\hspace*{-1.cm}
\includegraphics[width=.43\textwidth,angle =270]{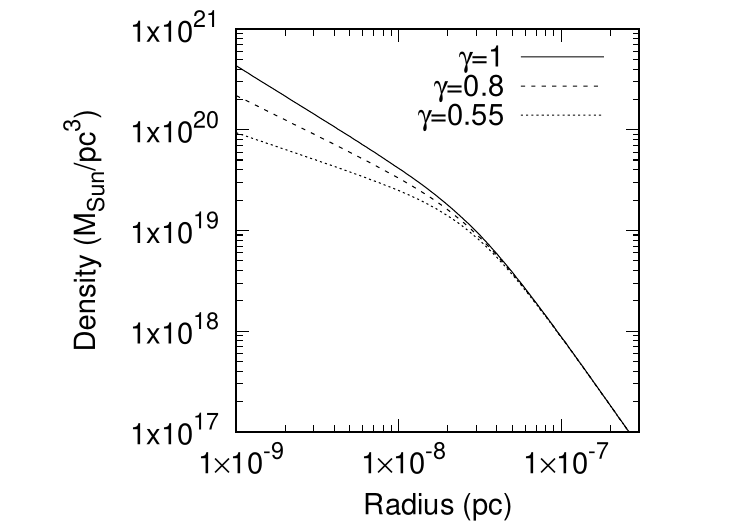}
\caption{   {
Dark matter spike density profile models corresponding to the density used in this work (Eq. \ref{dens}), for different values of inner power law slope $\gamma$. Solid line is for $\gamma=1$, dashed line for $\gamma=0.8$ and dotted line for $\gamma=0.55$. The models are normalized such to have the same density at infinity.}
}\label{models}
\end{figure}

\begin{figure*}
\includegraphics[width=1.\textwidth,angle =0]{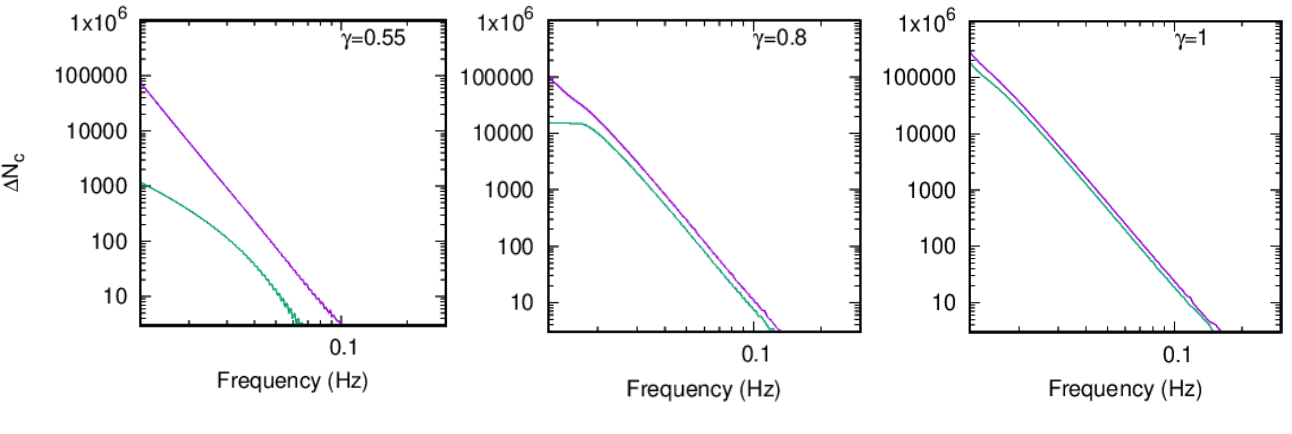}
\caption{
Change in the number of GW cycles with respect to the vacuum inspiral. Purple lines are obtained from equations \ref{eqa} and \ref{eqe} that include the contribution from the fast moving particles. Green lines were obtained using the standard Chandrasekhar's formula. 
}\label{Dcycles}
\end{figure*}

\section{Effect on the waveform: dephasing}\label{dephasing}

To quantify the size of the dephasing effect due to dynamical friction, we estimate the difference between the number of gravitational
wave cycles during the inspiral in vacuum and
in presence of a DM mini-spike, for various density profile models. We
calculate this for models where the contribution of fast moving particles is neglected as in the literature, and in models that include the effect of the fast particles.

We define the number of GW cycles by
integrating the GW frequency between two times
\begin{equation}
N_{\rm c}=\int_{t_1}^{t_2} f_{\rm GW}(t) dt\ .
\end{equation}
In the  approximation where the binary is circular, the GW frequency
 is twice the orbital frequency. 
 Eccentric binaries emit a GW signal with a broad spectrum of frequencies; the peak gravitational wave frequency corresponding to the harmonic which leads to the
maximal emission of GW radiation can be approximated
as \cite{2003ApJ...598..419W}
\begin{equation}
f_{\rm GW,peak}={\sqrt{GM}\over \pi}{1+e\over [a(1-e^2)]^{3/2}}\ ,
\end{equation}
{    and its time evolution is obtained from the evolution of $a$ and $e$ calculated from 
Eqs.~(\ref{eqa}) and~(\ref{eqe}).}

 The difference in the number of GW
cycles with and without DM, is then defined as
\begin{equation}
\Delta N_{\rm c}=N^{\rm vacum}_{\rm c}-N^{\rm DM}_{\rm c}\ .
\end{equation}
We assume that the central IMBH is surrounded by a DM
spike, formed as a consequence of the adiabatic growth of the central IMBH in a DM halo.
After its formation, the central density of the DM spike is likely to be lowered by several processes, which include the perturbative effect of inspiralling BHs \cite{2020PhRvD.102h3006K} and/or DM self-annihilation \cite{Bertone:2006nq,Bertone:2005hw,Bertone:2018xtm,2007PhRvD..76j3532V,2010arXiv1001.3706M}.    {We note that the  profile of the DM spike depends on the formation history of the central IMBH. If the IMBH has experienced disruptive processes such as mergers in the past, the mini-spike would be weakened or even disappear \cite{Eda:2014kra,Eda:2013gg}. Furthermore, self annihilations cause the DM density to decay which can result in a weak density plateau, $\propto r^{-0.5}$, near the SMBH \cite{Bertone:2006nq,Bertone:2005hw,Bertone:2018xtm,2007PhRvD..76j3532V,2010arXiv1001.3706M}. For the above reasons the slope of a power law DM spike profile $\rho_{\rm sp}(r) \propto r^{-\gamma}$ can be essentially treated  as a free parameter within the range $0 \lesssim \gamma \lesssim 3$ \cite{Eda:2014kra,Eda:2013gg}.} 

Correspondingly, we model the DM spike using a broken power law model:
\begin{equation}\label{dens}
\rho(r) =
\begin{cases}
    \rho_0 \left(  \frac{r}{r_0} \right)^{-\gamma}
    \left[
        1+\left( \frac{r}{r_0} \right)^{\alpha}
    \right]^{(\gamma-\gamma_{e})/{\alpha}}, & \text{if } r > r_{\text{in}} \\
    0, & \text{if } r \leq r_{\text{in}}
\end{cases}
\end{equation}
     {where $\rho_{0}$ is the density normalization, $r_{\rm in}=4GM_{\bullet}/c^{2}$ is the innermost stable circular orbit radius and  $\alpha$ is a parameter that defines the  transition strength between an inner power law cusp with slope $\gamma$ and the outer
 power law profile with slope $\gamma_e$. 
The scale $r_0$ is the radius where this transition occurs. }
 
The normalization of the density profile is chosen such that the density at infinity matches that of the density model in \cite{Eda:2014kra,Eda:2013gg}
\begin{equation}\label{den}
\rho_{\rm sp}(r)=\rho_{\rm sp}\left(  \frac{r}{r_{\rm sp}} \right)^{-\gamma_e}
\end{equation}
where
\begin{equation}
r_{\rm sp}\approx
\left[ 
\frac{(3-\gamma_e)0.2^{3-\gamma_e}M_{\bullet}}{2\pi \rho_{\rm sp}}\ 
\right]^{1/3}\ .
\end{equation}
and $\rho_{\rm sp}$ is the density normalization chosen to be $\rho_{\rm sp}=226 \ M_{\odot} \ \rm pc^{-3}$ \citep{2020PhRvD.102h3006K}.
   {
 This model is the fiducial model  investigated in \cite{Eda:2014kra,Eda:2013gg}. The slope $\gamma_{\rm e}=7/3$ is expected to develop in the center of a halo with an initial profile scaling  $\rho\sim r^{-1}$, such as an NFW profile. For this reason, we set $\gamma_{\rm e}=7/3$ and explore the dependence of our results on the assumed value of the inner slope $\gamma$, which we take to be $0.55$, $0.8$ and $1$.
 {    We set the break radius $r_0=3\times 10^{-8}\rm pc$, so that for the assumed initial galactocentric radius, $a_0=2\times 10^{-8}\rm pc$, the inspiral is  completely within the core, where  the dynamical friction effect is maximised. We set $\alpha=5$ that corresponds to a sharp transition. For any $\alpha \gtrsim 5$  our results remain essentially unchanged. However, as $\alpha$ is lowered the effect of the fast moving stars is somewhat reduced due to the models being more similar to the single power-law $\gamma_{e}=2.3$ model over a wider range of radii.}
 }

We show the density profile models in Fig\ \ref{models}.
We note that our choice of normalization is different with what often used in the literature. For example, \cite{2020PhRvD.102h3006K} use the model of Eq. \ref{den}, and vary $\gamma_{e}$ across a range of values. Given this choice, shallower profiles mean a much smaller central density. The result is that the effect of dynamical friction rapidly becomes unimportant for $\gamma_{e}\leq3/2$. Thus, the decreasing dynamical friction in this case is not because of the change in $\gamma_{e}$, but because of the different normalization that leads to much lower central densities.
On the other hand, our models all have the same normalization outside $r_0$, which keeps the density high inside this radius and dynamical friction important for most values of $\gamma$.

Generally, the distribution function $f(\upsilon_{\rm DM})$ corresponding to the density model in Eq. (\ref{dens}) and the potential generated by the DM and central IMBH cannot be obtained analytically. We therefore solve the Eddington equation numerically to obtain the distribution function and then compute numerically the integrals that appear in the RHS of equation\ (\ref{df}). We note that the first of these integrals can be simplified using the following expression  \citep{2006ApJ...648..890M,2012ApJ...745...83A}:
\begin{eqnarray}\label{fvr}
F(<v,r)=\int_{0}^{\upsilon}d\upsilon_{\rm DM} 4\pi f(\upsilon_{\rm DM})\upsilon_{\rm DM}^{2} = 1 - {1\over \rho}\int_{0}^E d\phi' {d\rho\over d\phi'} \nonumber \\
\times \left\{ 1 + {2\over\pi} \left[{v/\sqrt{2}\over\sqrt{\phi'-E}} - \tan^{-1}\left({v/\sqrt{2}\over\sqrt{\phi'-E}}\right)\right]\right\},
\end{eqnarray}
where $E=\frac{1}{2}v^2+\phi(r)$; $F(<v,r)$ is simply the fraction of DM particles at $r$ that move slower than the infalling BH.


Since we are interested to show a proof-of-concept example in this work, we simply consider a single set of initial conditions. Moreover, we only consider  circular orbits, and plan to look at eccentric orbits in a future work.
{    As stated above, the initial semi-major axis of the orbit is $a_0=2\times 10^{-8}\rm pc$, smaller than $r_0$.} 
Thus,  the binary effectively moves within a density spike of slope $\gamma$.  As before, the primary BH mass is $M_{\bullet}=1.4\times 10^{3}M_\odot$ and the binary mass ratio is 
$q=10^{-3}$. {   
Given these masses and the orbit, the binary GW frequency is
$8.9\times 10^{-3}Hz$; i.e., it is within the LISA frequency band $0.1\rm mHz\ - 1 \rm Hz$.  The orbit evolves completely within this frequency window 
until it reaches coalescence.
}

 The results of our calculation are shown in Fig.~\ref{Dcycles}. For $\gamma=0.55$, we see  that the dynamical friction contribution from the fast moving particles leads to nearly two orders of magnitude difference in the value of $\Delta N_c$ with respect to the standard treatment. The difference is larger at lower frequencies, but remains almost an order of magnitude throughout. As $\gamma$ is increased, the relative contribution of the fast moving particles to dynamical friction decreases. It is still important for $\gamma=0.8$ at $f\sim 10^{-2}$Hz, but at higher frequencies and/or for larger $\gamma$ the difference is small.

\section{conclusions}\label{conc}

In this work we have 
considered the evolution of a massive binary inside a 
DM density spike.
For the first time we included in the treatment of this problem, the dynamical friction 
produced by particles that move faster than the inspiraling BH, usually referred to as ``non-dominant'' term. 
This term is neglected in the standard Chandrasekhar treatment where all the frictional force is assumed to be produced by particles moving slower than the binary.
We have studied the effect of this term on the orbital decay, and on the circularization time of the binary. We then studied the dephasing of the gravitational waveform produced by the binary due to dynamical friction. Our main conclusions are summarized in what follows:
\begin{itemize}
\item[1] The evolution of the binary eccentricity due to dynamical friction is shown to be significantly affected. The parameter space where dynamical friction causes  the orbit to become more eccentric is enlarged when the contribution from the fast moving particles is included. In our treatment, dynamical friction leads the orbit to become more eccentric for any cusp with slope $\gamma< 1.8$, while the standard treatment would predict $\gamma< 1.5$ (Fig. 1). For $\gamma\gtrsim 1.8$, dynamical friction causes the orbit to
circularize faster than if evolved only due to energy
loss by GW radiation. For shallower slopes, instead, the orbit
is expected to circularize at a slower rate (Fig. 1).
\item[2] The timescale over which the binary eccentricity evolves is also significantly modified by the non-dominant terms. This statement appears to be true for most values of $\gamma$. For $\gamma=2$,  our treatment predicts a circularization timescale that is about 2 times shorter than the standard formula (Fig. 2).
\item[3]
For $\gamma\leq 1$ the orbital decay time of the binary due to dynamical friction is much shorter than predicted by the Chandrasekhar's formula. For $\gamma=1$ the difference is a factor of $2$. But,  as $\gamma$ approaches $0.5$, the error due to neglecting the fast moving particles becomes arbitrarily large (Fig. 2).

\item[4]
We calculate the dephasing of the GW signal due to dynamical friction.  We show  
that the dephasing of the gravitational waveform induced by DM can be much larger than previously
thought. 
The difference between the dephasing computed with the standard treatment and ours can be as large as two orders of magnitude for $\gamma\lesssim 0.6$, while it likely to be negligible  for any $\gamma\gtrsim 1$ (Fig. 4).
\end{itemize}

In this article we demonstrated that the dynamical friction  from the fast moving particles can have
a significant effect on the evolution of a massive binary within a DM spike.
The effect is very sensitive
to the slope of the DM distribution, rapidly becoming
important for $\gamma\lesssim 1$.
Shallow density cusps can be produced, for example, by the interaction of the inspiraling BH with the surrounding cusp, or by  DM self-annihilation. It is therefore our recommendation, that future similar studies will consider the replacement of the standard Chandrasekhar's formula with equations 
\ref{avinta} and \ref{avinte}.

Since we were interested in isolating the contribution of the fast moving DM particles to dynamical friction, we have based our models on a number of simplifying assumptions. We have assumed that the DM spike is not affected by the binary motion; we have assumed that the velocity distribution of the DM particles is isotropic; and we have ignored the possibility that DM is accreted directly onto the inspiraling BH. Moreover, our work does not take into account relativistic terms in the description of the orbital dynamics and
distribution of DM. Although  our assumptions are likely to break down in realistic situations, we expect that the fast moving particles will still play an important contribution to the dynamical friction force.


\begin{thebibliography}{35}%
\makeatletter
\providecommand \@ifxundefined [1]{%
 \@ifx{#1\undefined}
}%
\providecommand \@ifnum [1]{%
 \ifnum #1\expandafter \@firstoftwo
 \else \expandafter \@secondoftwo
 \fi
}%
\providecommand \@ifx [1]{%
 \ifx #1\expandafter \@firstoftwo
 \else \expandafter \@secondoftwo
 \fi
}%
\providecommand \natexlab [1]{#1}%
\providecommand \enquote  [1]{``#1''}%
\providecommand \bibnamefont  [1]{#1}%
\providecommand \bibfnamefont [1]{#1}%
\providecommand \citenamefont [1]{#1}%
\providecommand \href@noop [0]{\@secondoftwo}%
\providecommand \href [0]{\begingroup \@sanitize@url \@href}%
\providecommand \@href[1]{\@@startlink{#1}\@@href}%
\providecommand \@@href[1]{\endgroup#1\@@endlink}%
\providecommand \@sanitize@url [0]{\catcode `\\12\catcode `\$12\catcode
  `\&12\catcode `\#12\catcode `\^12\catcode `\_12\catcode `\%12\relax}%
\providecommand \@@startlink[1]{}%
\providecommand \@@endlink[0]{}%
\providecommand \url  [0]{\begingroup\@sanitize@url \@url }%
\providecommand \@url [1]{\endgroup\@href {#1}{\urlprefix }}%
\providecommand \urlprefix  [0]{URL }%
\providecommand \Eprint [0]{\href }%
\providecommand \doibase [0]{http://dx.doi.org/}%
\providecommand \selectlanguage [0]{\@gobble}%
\providecommand \bibinfo  [0]{\@secondoftwo}%
\providecommand \bibfield  [0]{\@secondoftwo}%
\providecommand \translation [1]{[#1]}%
\providecommand \BibitemOpen [0]{}%
\providecommand \bibitemStop [0]{}%
\providecommand \bibitemNoStop [0]{.\EOS\space}%
\providecommand \EOS [0]{\spacefactor3000\relax}%
\providecommand \BibitemShut  [1]{\csname bibitem#1\endcsname}%
\let\auto@bib@innerbib\@empty
\bibitem [{\citenamefont {Amaro-Seoane}\ \emph {et~al.}(2013)\citenamefont
  {Amaro-Seoane} \emph {et~al.}}]{AmaroSeoane:2012km}%
  \BibitemOpen
  \bibfield  {author} {\bibinfo {author} {\bibfnamefont {P.}~\bibnamefont
  {Amaro-Seoane}} \emph {et~al.},\ }\href@noop {} {\bibfield  {journal}
  {\bibinfo  {journal} {GW Notes}\ }    {\bibinfo {volume} {6}},\ \bibinfo
  {pages} {4} (\bibinfo {year} {2013})},\ \Eprint
  {http://arxiv.org/abs/1201.3621} {arXiv:1201.3621 [astro-ph.CO]} \BibitemShut
  {NoStop}%
\bibitem [{\citenamefont {{Amaro-Seoane}}\ \emph {et~al.}(2017)\citenamefont
  {{Amaro-Seoane}} \emph {et~al.}}]{2017arXiv170200786A}%
  \BibitemOpen
  \bibfield  {author} {\bibinfo {author} {\bibfnamefont {P.}~\bibnamefont
  {{Amaro-Seoane}}} \emph {et~al.},\ }\href@noop {} {\  (\bibinfo {year}
  {2017})},\ \Eprint {http://arxiv.org/abs/1702.00786} {arXiv:1702.00786
  [astro-ph.IM]} \BibitemShut {NoStop}%
\bibitem [{\citenamefont {Amaro-Seoane}\ \emph {et~al.}(2023)\citenamefont
  {Amaro-Seoane} \emph {et~al.}}]{2023LRR....26....2A}%
  \BibitemOpen
  \bibfield  {author} {\bibinfo {author} {\bibfnamefont {P.}~\bibnamefont
  {Amaro-Seoane}} \emph {et~al.},\ }\href {\doibase 10.1007/s41114-022-00041-y}
  {\bibfield  {journal} {\bibinfo  {journal} {Living Reviews in Relativity}\
  }    {\bibinfo {volume} {26}},\ \bibinfo {eid} {2} (\bibinfo {year}
  {2023})},\ \Eprint {http://arxiv.org/abs/2203.06016} {arXiv:2203.06016
  [gr-qc]} \BibitemShut {NoStop}%
\bibitem [{\citenamefont {Barausse}\ \emph {et~al.}(2014)\citenamefont
  {Barausse}, \citenamefont {Cardoso},\ and\ \citenamefont
  {Pani}}]{Barausse:2014tra}%
  \BibitemOpen
  \bibfield  {author} {\bibinfo {author} {\bibfnamefont {E.}~\bibnamefont
  {Barausse}}, \bibinfo {author} {\bibfnamefont {V.}~\bibnamefont {Cardoso}}, \
  and\ \bibinfo {author} {\bibfnamefont {P.}~\bibnamefont {Pani}},\ }\href
  {\doibase 10.1103/PhysRevD.89.104059} {\bibfield  {journal} {\bibinfo
  {journal} {Phys. Rev.}\ }    {\bibinfo {volume} {D89}},\ \bibinfo {pages}
  {104059} (\bibinfo {year} {2014})},\ \Eprint {http://arxiv.org/abs/1404.7149}
  {arXiv:1404.7149 [gr-qc]} \BibitemShut {NoStop}%
\bibitem [{\citenamefont {Gondolo}\ and\ \citenamefont
  {Silk}(1999)}]{Gondolo:1999ef}%
  \BibitemOpen
  \bibfield  {author} {\bibinfo {author} {\bibfnamefont {P.}~\bibnamefont
  {Gondolo}}\ and\ \bibinfo {author} {\bibfnamefont {J.}~\bibnamefont {Silk}},\
  }\href {\doibase 10.1103/PhysRevLett.83.1719} {\bibfield  {journal} {\bibinfo
   {journal} {Phys. Rev. Lett.}\ }    {\bibinfo {volume} {83}},\ \bibinfo
  {pages} {1719} (\bibinfo {year} {1999})},\ \Eprint
  {http://arxiv.org/abs/astro-ph/9906391} {arXiv:astro-ph/9906391 [astro-ph]}
  \BibitemShut {NoStop}%
\bibitem [{\citenamefont {Sadeghian}\ \emph {et~al.}(2013)\citenamefont
  {Sadeghian}, \citenamefont {Ferrer},\ and\ \citenamefont
  {Will}}]{Sadeghian:2013laa}%
  \BibitemOpen
  \bibfield  {author} {\bibinfo {author} {\bibfnamefont {L.}~\bibnamefont
  {Sadeghian}}, \bibinfo {author} {\bibfnamefont {F.}~\bibnamefont {Ferrer}}, \
  and\ \bibinfo {author} {\bibfnamefont {C.~M.}\ \bibnamefont {Will}},\ }\href
  {\doibase 10.1103/PhysRevD.88.063522} {\bibfield  {journal} {\bibinfo
  {journal} {Phys. Rev.}\ }    {\bibinfo {volume} {D88}},\ \bibinfo {pages}
  {063522} (\bibinfo {year} {2013})},\ \Eprint {http://arxiv.org/abs/1305.2619}
  {arXiv:1305.2619 [astro-ph.GA]} \BibitemShut {NoStop}%
\bibitem [{\citenamefont {Eda}\ \emph {et~al.}(2013)\citenamefont {Eda},
  \citenamefont {Itoh}, \citenamefont {Kuroyanagi},\ and\ \citenamefont
  {Silk}}]{Eda:2013gg}%
  \BibitemOpen
  \bibfield  {author} {\bibinfo {author} {\bibfnamefont {K.}~\bibnamefont
  {Eda}}, \bibinfo {author} {\bibfnamefont {Y.}~\bibnamefont {Itoh}}, \bibinfo
  {author} {\bibfnamefont {S.}~\bibnamefont {Kuroyanagi}}, \ and\ \bibinfo
  {author} {\bibfnamefont {J.}~\bibnamefont {Silk}},\ }\href {\doibase
  10.1103/PhysRevLett.110.221101} {\bibfield  {journal} {\bibinfo  {journal}
  {Phys. Rev. Lett.}\ }    {\bibinfo {volume} {110}},\ \bibinfo {pages}
  {221101} (\bibinfo {year} {2013})},\ \Eprint {http://arxiv.org/abs/1301.5971}
  {arXiv:1301.5971 [gr-qc]} \BibitemShut {NoStop}%
\bibitem [{\citenamefont {Eda}\ \emph {et~al.}(2015)\citenamefont {Eda},
  \citenamefont {Itoh}, \citenamefont {Kuroyanagi},\ and\ \citenamefont
  {Silk}}]{Eda:2014kra}%
  \BibitemOpen
  \bibfield  {author} {\bibinfo {author} {\bibfnamefont {K.}~\bibnamefont
  {Eda}}, \bibinfo {author} {\bibfnamefont {Y.}~\bibnamefont {Itoh}}, \bibinfo
  {author} {\bibfnamefont {S.}~\bibnamefont {Kuroyanagi}}, \ and\ \bibinfo
  {author} {\bibfnamefont {J.}~\bibnamefont {Silk}},\ }\href {\doibase
  10.1103/PhysRevD.91.044045} {\bibfield  {journal} {\bibinfo  {journal} {Phys.
  Rev.}\ }    {\bibinfo {volume} {D91}},\ \bibinfo {pages} {044045}
  (\bibinfo {year} {2015})},\ \Eprint {http://arxiv.org/abs/1408.3534}
  {arXiv:1408.3534 [gr-qc]} \BibitemShut {NoStop}%
\bibitem [{\citenamefont {Yue}\ and\ \citenamefont {Han}(2018)}]{Yue:2017iwc}%
  \BibitemOpen
  \bibfield  {author} {\bibinfo {author} {\bibfnamefont {X.-J.}\ \bibnamefont
  {Yue}}\ and\ \bibinfo {author} {\bibfnamefont {W.-B.}\ \bibnamefont {Han}},\
  }\href {\doibase 10.1103/PhysRevD.97.064003} {\bibfield  {journal} {\bibinfo
  {journal} {Phys. Rev.}\ }    {\bibinfo {volume} {D97}},\ \bibinfo {pages}
  {064003} (\bibinfo {year} {2018})},\ \Eprint
  {http://arxiv.org/abs/1711.09706} {arXiv:1711.09706 [gr-qc]} \BibitemShut
  {NoStop}%
\bibitem [{\citenamefont {Macedo}\ \emph {et~al.}(2013)\citenamefont {Macedo},
  \citenamefont {Pani}, \citenamefont {Cardoso},\ and\ \citenamefont
  {Crispino}}]{Macedo:2013qea}%
  \BibitemOpen
  \bibfield  {author} {\bibinfo {author} {\bibfnamefont {C.~F.~B.}\
  \bibnamefont {Macedo}}, \bibinfo {author} {\bibfnamefont {P.}~\bibnamefont
  {Pani}}, \bibinfo {author} {\bibfnamefont {V.}~\bibnamefont {Cardoso}}, \
  and\ \bibinfo {author} {\bibfnamefont {L.~C.~B.}\ \bibnamefont {Crispino}},\
  }\href {\doibase 10.1088/0004-637X/774/1/48} {\bibfield  {journal} {\bibinfo
  {journal} {Astrophys. J.}\ }    {\bibinfo {volume} {774}},\ \bibinfo
  {pages} {48} (\bibinfo {year} {2013})},\ \Eprint
  {http://arxiv.org/abs/1302.2646} {arXiv:1302.2646 [gr-qc]} \BibitemShut
  {NoStop}%
\bibitem [{\citenamefont {Yue}\ \emph {et~al.}(2019)\citenamefont {Yue},
  \citenamefont {Han},\ and\ \citenamefont {Chen}}]{Yue:2019ndw}%
  \BibitemOpen
  \bibfield  {author} {\bibinfo {author} {\bibfnamefont {X.-J.}\ \bibnamefont
  {Yue}}, \bibinfo {author} {\bibfnamefont {W.-B.}\ \bibnamefont {Han}}, \ and\
  \bibinfo {author} {\bibfnamefont {X.}~\bibnamefont {Chen}},\ }\href {\doibase
  10.3847/1538-4357/ab06f6} {\bibfield  {journal} {\bibinfo  {journal}
  {Astrophys. J.}\ }    {\bibinfo {volume} {874}},\ \bibinfo {pages} {34}
  (\bibinfo {year} {2019})},\ \Eprint {http://arxiv.org/abs/1802.03739}
  {arXiv:1802.03739 [gr-qc]} \BibitemShut {NoStop}%
\bibitem [{\citenamefont {Cardoso}\ and\ \citenamefont
  {Maselli}(2019)}]{Cardoso:2019rou}%
  \BibitemOpen
  \bibfield  {author} {\bibinfo {author} {\bibfnamefont {V.}~\bibnamefont
  {Cardoso}}\ and\ \bibinfo {author} {\bibfnamefont {A.}~\bibnamefont
  {Maselli}},\ }\href@noop {} {\  (\bibinfo {year} {2019})},\ \Eprint
  {http://arxiv.org/abs/1909.05870} {arXiv:1909.05870 [astro-ph.HE]}
  \BibitemShut {NoStop}%
\bibitem [{\citenamefont {{Kavanagh}}\ \emph {et~al.}(2020)\citenamefont
  {{Kavanagh}}, \citenamefont {{Nichols}}, \citenamefont {{Bertone}},\ and\
  \citenamefont {{Gaggero}}}]{2020PhRvD.102h3006K}%
  \BibitemOpen
  \bibfield  {author} {\bibinfo {author} {\bibfnamefont {B.~J.}\ \bibnamefont
  {{Kavanagh}}}, \bibinfo {author} {\bibfnamefont {D.~A.}\ \bibnamefont
  {{Nichols}}}, \bibinfo {author} {\bibfnamefont {G.}~\bibnamefont
  {{Bertone}}}, \ and\ \bibinfo {author} {\bibfnamefont {D.}~\bibnamefont
  {{Gaggero}}},\ }\href {\doibase 10.1103/PhysRevD.102.083006} {\bibfield
  {journal} {\bibinfo  {journal} {\prd}\ }    {\bibinfo {volume} {102}},\
  \bibinfo {eid} {083006} (\bibinfo {year} {2020})},\ \Eprint
  {http://arxiv.org/abs/2002.12811} {arXiv:2002.12811 [gr-qc]} \BibitemShut
  {NoStop}%
\bibitem [{\citenamefont {{Becker}}\ and\ \citenamefont
  {{Sagunski}}(2023)}]{2023PhRvD.107h3003B}%
  \BibitemOpen
  \bibfield  {author} {\bibinfo {author} {\bibfnamefont {N.}~\bibnamefont
  {{Becker}}}\ and\ \bibinfo {author} {\bibfnamefont {L.}~\bibnamefont
  {{Sagunski}}},\ }\href {\doibase 10.1103/PhysRevD.107.083003} {\bibfield
  {journal} {\bibinfo  {journal} {\prd}\ }    {\bibinfo {volume} {107}},\
  \bibinfo {eid} {083003} (\bibinfo {year} {2023})},\ \Eprint
  {http://arxiv.org/abs/2211.05145} {arXiv:2211.05145 [gr-qc]} \BibitemShut
  {NoStop}%
\bibitem [{\citenamefont
  {Chandrasekhar}(1943{\natexlab{a}})}]{Chandrasekhar1943a}%
  \BibitemOpen
  \bibfield  {author} {\bibinfo {author} {\bibfnamefont {S.}~\bibnamefont
  {Chandrasekhar}},\ }\href {\doibase 10.1086/144517} {\bibfield  {journal}
  {\bibinfo  {journal} {The Astrophysical Journal}\ }    {\bibinfo {volume}
  {97}},\ \bibinfo {pages} {255} (\bibinfo {year}
  {1943}{\natexlab{a}})}\BibitemShut {NoStop}%
\bibitem [{\citenamefont
  {Chandrasekhar}(1943{\natexlab{b}})}]{Chandrasekhar1943b}%
  \BibitemOpen
  \bibfield  {author} {\bibinfo {author} {\bibfnamefont {S.}~\bibnamefont
  {Chandrasekhar}},\ }\href {\doibase 10.1086/144518} {\bibfield  {journal}
  {\bibinfo  {journal} {The Astrophysical Journal}\ }    {\bibinfo {volume}
  {97}},\ \bibinfo {pages} {263} (\bibinfo {year}
  {1943}{\natexlab{b}})}\BibitemShut {NoStop}%
\bibitem [{\citenamefont {Merritt}\ \emph {et~al.}(2007)\citenamefont
  {Merritt}, \citenamefont {Harfst},\ and\ \citenamefont
  {Bertone}}]{Merritt:2006mt}%
  \BibitemOpen
  \bibfield  {author} {\bibinfo {author} {\bibfnamefont {D.}~\bibnamefont
  {Merritt}}, \bibinfo {author} {\bibfnamefont {S.}~\bibnamefont {Harfst}}, \
  and\ \bibinfo {author} {\bibfnamefont {G.}~\bibnamefont {Bertone}},\ }\href
  {\doibase 10.1103/PhysRevD.75.043517} {\bibfield  {journal} {\bibinfo
  {journal} {Phys. Rev.}\ }    {\bibinfo {volume} {D75}},\ \bibinfo {pages}
  {043517} (\bibinfo {year} {2007})},\ \Eprint
  {http://arxiv.org/abs/astro-ph/0610425} {arXiv:astro-ph/0610425 [astro-ph]}
  \BibitemShut {NoStop}%
\bibitem [{\citenamefont {{Dosopoulou}}\ \emph {et~al.}(2021)\citenamefont
  {{Dosopoulou}}, \citenamefont {{Greene}},\ and\ \citenamefont
  {{Ma}}}]{2021ApJ...922...40D}%
  \BibitemOpen
  \bibfield  {author} {\bibinfo {author} {\bibfnamefont {F.}~\bibnamefont
  {{Dosopoulou}}}, \bibinfo {author} {\bibfnamefont {J.~E.}\ \bibnamefont
  {{Greene}}}, \ and\ \bibinfo {author} {\bibfnamefont {C.-P.}\ \bibnamefont
  {{Ma}}},\ }\href {\doibase 10.3847/1538-4357/ac1fe4} {\bibfield  {journal}
  {\bibinfo  {journal} {\apj}\ }    {\bibinfo {volume} {922}},\ \bibinfo
  {eid} {40} (\bibinfo {year} {2021})},\ \Eprint
  {http://arxiv.org/abs/2108.08317} {arXiv:2108.08317 [astro-ph.GA]}
  \BibitemShut {NoStop}%
\bibitem [{\citenamefont {{Speeney}}\ \emph {et~al.}(2022)\citenamefont
  {{Speeney}}, \citenamefont {{Antonelli}}, \citenamefont {{Baibhav}},\ and\
  \citenamefont {{Berti}}}]{2022PhRvD.106d4027S}%
  \BibitemOpen
  \bibfield  {author} {\bibinfo {author} {\bibfnamefont {N.}~\bibnamefont
  {{Speeney}}}, \bibinfo {author} {\bibfnamefont {A.}~\bibnamefont
  {{Antonelli}}}, \bibinfo {author} {\bibfnamefont {V.}~\bibnamefont
  {{Baibhav}}}, \ and\ \bibinfo {author} {\bibfnamefont {E.}~\bibnamefont
  {{Berti}}},\ }\href {\doibase 10.1103/PhysRevD.106.044027} {\bibfield
  {journal} {\bibinfo  {journal} {\prd}\ }    {\bibinfo {volume} {106}},\
  \bibinfo {eid} {044027} (\bibinfo {year} {2022})},\ \Eprint
  {http://arxiv.org/abs/2204.12508} {arXiv:2204.12508 [gr-qc]} \BibitemShut
  {NoStop}%
\bibitem [{\citenamefont {{Chiari}}\ and\ \citenamefont {{Di
  Cintio}}(2022)}]{2022arXiv220705728C}%
  \BibitemOpen
  \bibfield  {author} {\bibinfo {author} {\bibfnamefont {C.}~\bibnamefont
  {{Chiari}}}\ and\ \bibinfo {author} {\bibfnamefont {P.}~\bibnamefont {{Di
  Cintio}}},\ }\href {\doibase 10.48550/arXiv.2207.05728} {\bibfield  {journal}
  {\bibinfo  {journal} {arXiv e-prints}\ ,\ \bibinfo {eid} {arXiv:2207.05728}}
  (\bibinfo {year} {2022})},\ \Eprint {http://arxiv.org/abs/2207.05728}
  {arXiv:2207.05728 [astro-ph.GA]} \BibitemShut {NoStop}%
\bibitem [{\citenamefont {{Binney}}\ and\ \citenamefont
  {{Tremaine}}(2008)}]{BinneyAndTremaine}%
  \BibitemOpen
  \bibfield  {author} {\bibinfo {author} {\bibfnamefont {J.}~\bibnamefont
  {{Binney}}}\ and\ \bibinfo {author} {\bibfnamefont {S.}~\bibnamefont
  {{Tremaine}}},\ }\href@noop {} {\emph {\bibinfo {title} {Galactic Dynamics:
  Second Edition, by James Binney and Scott Tremaine.~ISBN 978-0-691-13026-2
  (HB).~Published by Princeton University Press, Princeton, NJ USA, 2008.}}}\
  (\bibinfo  {publisher} {Princeton University Press},\ \bibinfo {year}
  {2008})\BibitemShut {NoStop}%
\bibitem [{\citenamefont {{Antonini}}\ and\ \citenamefont
  {{Merritt}}(2012)}]{2012ApJ...745...83A}%
  \BibitemOpen
  \bibfield  {author} {\bibinfo {author} {\bibfnamefont {F.}~\bibnamefont
  {{Antonini}}}\ and\ \bibinfo {author} {\bibfnamefont {D.}~\bibnamefont
  {{Merritt}}},\ }\href {\doibase 10.1088/0004-637X/745/1/83} {\bibfield
  {journal} {\bibinfo  {journal} {\apj}\ }    {\bibinfo {volume} {745}},\
  \bibinfo {eid} {83} (\bibinfo {year} {2012})},\ \Eprint
  {http://arxiv.org/abs/1108.1163} {arXiv:1108.1163 [astro-ph.GA]} \BibitemShut
  {NoStop}%
\bibitem [{\citenamefont {{Dosopoulou}}\ and\ \citenamefont
  {{Antonini}}(2017)}]{2017ApJ...840...31D}%
  \BibitemOpen
  \bibfield  {author} {\bibinfo {author} {\bibfnamefont {F.}~\bibnamefont
  {{Dosopoulou}}}\ and\ \bibinfo {author} {\bibfnamefont {F.}~\bibnamefont
  {{Antonini}}},\ }\href {\doibase 10.3847/1538-4357/aa6b58} {\bibfield
  {journal} {\bibinfo  {journal} {\apj}\ }    {\bibinfo {volume} {840}},\
  \bibinfo {eid} {31} (\bibinfo {year} {2017})},\ \Eprint
  {http://arxiv.org/abs/1611.06573} {arXiv:1611.06573 [astro-ph.GA]}
  \BibitemShut {NoStop}%
\bibitem [{\citenamefont {{Peters}}(1964)}]{1964PhRv..136.1224P}%
  \BibitemOpen
  \bibfield  {author} {\bibinfo {author} {\bibfnamefont {P.~C.}\ \bibnamefont
  {{Peters}}},\ }\href {\doibase 10.1103/PhysRev.136.B1224} {\bibfield
  {journal} {\bibinfo  {journal} {Physical Review}\ }    {\bibinfo {volume}
  {136}},\ \bibinfo {pages} {1224} (\bibinfo {year} {1964})}\BibitemShut
  {NoStop}%
\bibitem [{\citenamefont {{Merritt}}(2013)}]{2013degn.book.....M}%
  \BibitemOpen
  \bibfield  {author} {\bibinfo {author} {\bibfnamefont {D.}~\bibnamefont
  {{Merritt}}},\ }\href@noop {} {\emph {\bibinfo {title} {{Dynamics and
  Evolution of Galactic Nuclei}}}}\ (\bibinfo {year} {2013})\BibitemShut
  {NoStop}%
\bibitem [{\citenamefont {{Cardoso}}\ \emph {et~al.}(2021)\citenamefont
  {{Cardoso}}, \citenamefont {{Macedo}},\ and\ \citenamefont
  {{Vicente}}}]{2021PhRvD.103b3015C}%
  \BibitemOpen
  \bibfield  {author} {\bibinfo {author} {\bibfnamefont {V.}~\bibnamefont
  {{Cardoso}}}, \bibinfo {author} {\bibfnamefont {C.~F.~B.}\ \bibnamefont
  {{Macedo}}}, \ and\ \bibinfo {author} {\bibfnamefont {R.}~\bibnamefont
  {{Vicente}}},\ }\href {\doibase 10.1103/PhysRevD.103.023015} {\bibfield
  {journal} {\bibinfo  {journal} {\prd}\ }    {\bibinfo {volume} {103}},\
  \bibinfo {eid} {023015} (\bibinfo {year} {2021})},\ \Eprint
  {http://arxiv.org/abs/2010.15151} {arXiv:2010.15151 [gr-qc]} \BibitemShut
  {NoStop}%
\bibitem [{\citenamefont {{Becker}}\ \emph {et~al.}(2022)\citenamefont
  {{Becker}}, \citenamefont {{Sagunski}}, \citenamefont {{Prinz}},\ and\
  \citenamefont {{Rastgoo}}}]{2022PhRvD.105f3029B}%
  \BibitemOpen
  \bibfield  {author} {\bibinfo {author} {\bibfnamefont {N.}~\bibnamefont
  {{Becker}}}, \bibinfo {author} {\bibfnamefont {L.}~\bibnamefont
  {{Sagunski}}}, \bibinfo {author} {\bibfnamefont {L.}~\bibnamefont {{Prinz}}},
  \ and\ \bibinfo {author} {\bibfnamefont {S.}~\bibnamefont {{Rastgoo}}},\
  }\href {\doibase 10.1103/PhysRevD.105.063029} {\bibfield  {journal} {\bibinfo
   {journal} {\prd}\ }    {\bibinfo {volume} {105}},\ \bibinfo {eid}
  {063029} (\bibinfo {year} {2022})},\ \Eprint
  {http://arxiv.org/abs/2112.09586} {arXiv:2112.09586 [gr-qc]} \BibitemShut
  {NoStop}%
\bibitem [{\citenamefont {{Gould}}\ and\ \citenamefont
  {{Quillen}}(2003)}]{2003ApJ...592..935G}%
  \BibitemOpen
  \bibfield  {author} {\bibinfo {author} {\bibfnamefont {A.}~\bibnamefont
  {{Gould}}}\ and\ \bibinfo {author} {\bibfnamefont {A.~C.}\ \bibnamefont
  {{Quillen}}},\ }\href {\doibase 10.1086/375840} {\bibfield  {journal}
  {\bibinfo  {journal} {\apj}\ }    {\bibinfo {volume} {592}},\ \bibinfo
  {pages} {935} (\bibinfo {year} {2003})},\ \Eprint
  {http://arxiv.org/abs/astro-ph/0302437} {arXiv:astro-ph/0302437 [astro-ph]}
  \BibitemShut {NoStop}%
\bibitem [{\citenamefont {{Wen}}(2003)}]{2003ApJ...598..419W}%
  \BibitemOpen
  \bibfield  {author} {\bibinfo {author} {\bibfnamefont {L.}~\bibnamefont
  {{Wen}}},\ }\href {\doibase 10.1086/378794} {\bibfield  {journal} {\bibinfo
  {journal} {\apj}\ }    {\bibinfo {volume} {598}},\ \bibinfo {pages} {419}
  (\bibinfo {year} {2003})},\ \Eprint {http://arxiv.org/abs/astro-ph/0211492}
  {arXiv:astro-ph/0211492 [astro-ph]} \BibitemShut {NoStop}%
\bibitem [{\citenamefont {Bertone}(2006)}]{Bertone:2006nq}%
  \BibitemOpen
  \bibfield  {author} {\bibinfo {author} {\bibfnamefont {G.}~\bibnamefont
  {Bertone}},\ }\href {\doibase 10.1103/PhysRevD.73.103519} {\bibfield
  {journal} {\bibinfo  {journal} {Phys. Rev.}\ }    {\bibinfo {volume}
  {D73}},\ \bibinfo {pages} {103519} (\bibinfo {year} {2006})},\ \Eprint
  {http://arxiv.org/abs/astro-ph/0603148} {arXiv:astro-ph/0603148 [astro-ph]}
  \BibitemShut {NoStop}%
\bibitem [{\citenamefont {Bertone}\ and\ \citenamefont
  {Merritt}(2005)}]{Bertone:2005hw}%
  \BibitemOpen
  \bibfield  {author} {\bibinfo {author} {\bibfnamefont {G.}~\bibnamefont
  {Bertone}}\ and\ \bibinfo {author} {\bibfnamefont {D.}~\bibnamefont
  {Merritt}},\ }\href {\doibase 10.1103/PhysRevD.72.103502} {\bibfield
  {journal} {\bibinfo  {journal} {Phys. Rev.}\ }    {\bibinfo {volume}
  {D72}},\ \bibinfo {pages} {103502} (\bibinfo {year} {2005})},\ \Eprint
  {http://arxiv.org/abs/astro-ph/0501555} {arXiv:astro-ph/0501555 [astro-ph]}
  \BibitemShut {NoStop}%
\bibitem [{\citenamefont {Bertone}\ and\ \citenamefont
  {Tait}(2018)}]{Bertone:2018xtm}%
  \BibitemOpen
  \bibfield  {author} {\bibinfo {author} {\bibfnamefont {G.}~\bibnamefont
  {Bertone}}\ and\ \bibinfo {author} {\bibfnamefont {M.~P.}\ \bibnamefont
  {Tait}, \bibfnamefont {Tim}},\ }\href {\doibase 10.1038/s41586-018-0542-z}
  {\bibfield  {journal} {\bibinfo  {journal} {Nature}\ }    {\bibinfo
  {volume} {562}},\ \bibinfo {pages} {51} (\bibinfo {year} {2018})},\ \Eprint
  {http://arxiv.org/abs/1810.01668} {arXiv:1810.01668 [astro-ph.CO]}
  \BibitemShut {NoStop}%
\bibitem [{\citenamefont {{Vasiliev}}(2007)}]{2007PhRvD..76j3532V}%
  \BibitemOpen
  \bibfield  {author} {\bibinfo {author} {\bibfnamefont {E.}~\bibnamefont
  {{Vasiliev}}},\ }\href {\doibase 10.1103/PhysRevD.76.103532} {\bibfield
  {journal} {\bibinfo  {journal} {\prd}\ }    {\bibinfo {volume} {76}},\
  \bibinfo {eid} {103532} (\bibinfo {year} {2007})},\ \Eprint
  {http://arxiv.org/abs/0707.3334} {arXiv:0707.3334 [astro-ph]} \BibitemShut
  {NoStop}%
\bibitem [{\citenamefont {{Merritt}}(2010)}]{2010arXiv1001.3706M}%
  \BibitemOpen
  \bibfield  {author} {\bibinfo {author} {\bibfnamefont {D.}~\bibnamefont
  {{Merritt}}},\ }\href {\doibase 10.48550/arXiv.1001.3706} {\bibfield
  {journal} {\bibinfo  {journal} {arXiv e-prints}\ ,\ \bibinfo {eid}
  {arXiv:1001.3706}} (\bibinfo {year} {2010})},\ \Eprint
  {http://arxiv.org/abs/1001.3706} {arXiv:1001.3706 [astro-ph.CO]} \BibitemShut
  {NoStop}%
\bibitem [{\citenamefont {{Merritt}}\ and\ \citenamefont
  {{Szell}}(2006)}]{2006ApJ...648..890M}%
  \BibitemOpen
  \bibfield  {author} {\bibinfo {author} {\bibfnamefont {D.}~\bibnamefont
  {{Merritt}}}\ and\ \bibinfo {author} {\bibfnamefont {A.}~\bibnamefont
  {{Szell}}},\ }\href {\doibase 10.1086/506010} {\bibfield  {journal} {\bibinfo
   {journal} {\apj}\ }    {\bibinfo {volume} {648}},\ \bibinfo {pages}
  {890} (\bibinfo {year} {2006})},\ \Eprint
  {http://arxiv.org/abs/astro-ph/0510498} {arXiv:astro-ph/0510498 [astro-ph]}
  \BibitemShut {NoStop}%
\end{thebibliography}
\end{document}